\documentclass[aps,prl,reprint,showpacs,amsmath,amssymb,10pt]{revtex4-1} % PRL -superscriptaddress

\usepackage{graphicx}       % Include figure files
\usepackage{txfonts}        % Times New Roman text & math font
\usepackage[plainpages=false,pdfpagelabels,colorlinks]{hyperref} % this line at end of preamble
\hypersetup{colorlinks,citecolor=blue,filecolor=blue,linkcolor=blue,urlcolor=blue}

\newcommand{\gvec}[1]{\boldsymbol{\mathrm{#1}}}			        % my vector style
\newcommand{\ket}[1]{\left|\hspace{0.5pt} #1 \right\rangle}	% my ket style
\newcommand{\sket}[1]{|\hspace{0.5pt} #1 \rangle}	          % my small ket style

\begin{document}

%\preprint{APS/123-QED}

\title{Texture control in a pseudospin Bose-Einstein condensate}

\author{Gary Ruben}%
 \email{gary.ruben@monash.edu}
\author{Michael J. Morgan}
\author{David M. Paganin}
\affiliation{School of Physics, Monash University, Victoria 3800, Australia}

\date{May 21, 2010}
%\date{\today}

\begin{abstract}

We describe a wavefunction engineering approach to the formation of textures in a two-component nonrotated Bose-Einstein condensate. By controlling the phases of wavepackets that combine in a three-wave interference process, a ballistically-expanding regular lattice-texture is generated, in which the phases determine the component textures.
A particular example is presented of a lattice-texture composed of half-quantum vortices and spin-2 textures. We demonstrate the lattice formation with numerical simulations of a viable experiment, identifying the textures and relating their locations to a linear theory of wavepacket interference.

\end{abstract}

\pacs{03.75.Mn, 03.75.Lm, 03.75.Dg}      % PACS codes
% PACS code candidates:\\
% 02.40.Xx Singularity theory\\
% 03.65.Vf Geometric/Topological phases (quantum mechanics)\\
% 03.75.Dg Atom and neutron interferometry\\
% 03.75.Lm Vortices in BECs\\
% 03.75.Mn Multicomponent condensates; spinor condensates 
% 03.75.Kk dynamic properties\\
% 07.60.Ly Interferometers\\
% 47.32.C- Vortex dynamics (fluid flow)\\
% 47.32.Cc Vorticity in rotational flows (fluid)\\

%\keywords{Suggested keywords} %Use showkeys class option if keyword display desired

\maketitle

Topological spin textures arise in magnetic materials \cite{MBi09}, in director fields of liquid crystals \cite{No97}, in field theoretic models of particles \cite{Sky61}, and in models of the early universe \cite{Vil81}.
Multicomponent Bose-Einstein condensates (BECs) may act as analogues of these and other condensed matter systems, enabling the study of phenomena that may be otherwise inaccessible to experimental investigation.

Bulk rotation of scalar (single-component) BECs provides one method for creating a regular lattice of quantized vortices, associated with mass currents. The additional spin freedom in rotating multicomponent BECs suggest the existence of related spin lattice-textures, which are indeed observed \cite{MizKo04, *ReiVa04, *Mue04}.

In this Letter we describe a complementary method for the controlled production of a lattice-texture in a multicomponent BEC that forgoes bulk rotation of the condensate by instead exploiting interference to produce expanding lattices of singly-quantized vortices in any or all components. An advantage of this method is that the textures comprising the motif may be directly determined by the vortex-lattice alignment, which depends on the controllable wavepacket phases. Experimental production of isolated textures has been demonstrated within pseudospin-$\tfrac12$, spin-1, and spin-2 BECs \cite{MatAn99, LeaSh03, LesHa09}. Textures associated with small numbers of vortices in the lattice-texture could subsequently be isolated from the dynamically-expanding lattice by retrapping them optically.

\begin{figure}[b!]
%    \centering
%    \includegraphics[width=79mm]{fig1}
    \includegraphics{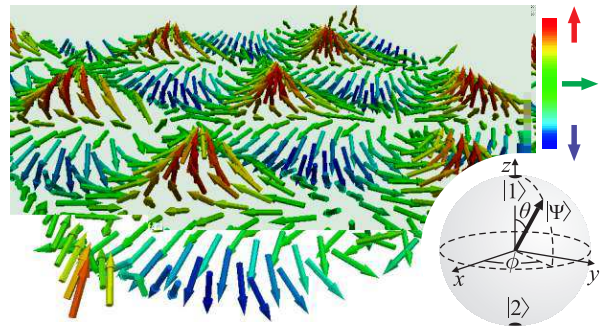}
  \caption{(color online) A planar lattice texture is created by the interference of a pseudospin-$\tfrac12$ BEC initially separated into three pieces. The hexagonal lattice, visualized with Bloch vectors in this numerical simulation, has a motif containing three textures: one half-quantum vortex of each sign and a spin-2 texture.}
  \label{fig:carpet_with_bloch}
\end{figure}

Nonrotated single-component BECs accommodate regular vortex lattices, created by a three-wave linear interference process \cite{RubPa08}. In this scenario, three initially-separated BEC wavepackets expand and interfere, in an analogous process to a Young's three-pinhole interferometer \cite{RubPa07, *RubPa07a}. When the initial wavepackets are arranged at the corners of an equilateral triangle, the resulting lattice has honeycomb symmetry and can be thought of as a dynamically-expanding Abrikosov lattice (a hexagonal lattice containing unit-charge vortices of one circulation), interleaved with a second similar lattice of vortices with opposite circulation.
It is probable that such honeycomb vortex-antivortex (VA) lattices have been experimentally generated in BECs, although this has neither been recognized nor verified directly---see \cite{SchWe07, *CarAn08, HenRy09}.
The initial wavepacket phases establish particular $xy$-translations of the resulting 2D VA lattices \cite{RubPa08}. By engineering the initial wavepackets in a two-component BEC, we exploit this phase dependence to produce VA lattices within each component of a two-component BEC, aligning them to form the dynamically-expanding lattice texture. In the example in Fig.~\ref{fig:carpet_with_bloch}, the position-dependent state is represented by local Bloch vectors, which project the state onto the surface of a unit-radius Bloch sphere (Fig.~\ref{fig:carpet_with_bloch} inset).

Although a rotating BEC is extended in three spatial dimensions, its resident vortex lattice is primarily 2D in nature, which restricts the topology of any associated texture. The VA lattice is similarly 2D in nature, because the three initial wavepacket locations together define a plane. In both cases, axial absorption imaging captures the 2D structure by projecting the density parallel to the vortex core nodal-lines. The simplest texture arises when a vortex in one component coincides with a vortex-free region in the second component to create a half-quantum vortex (HQV) \cite{CroBr77}, which is a counterpart to the Alice string in particle physics \cite{LeoVo00, *IsoMa01}; its detection in ${}^3$He-$A$ is sought as evidence of a spin triplet superconducting state in Sr${}_2$RuO${_4}$ \cite{VakLe09}. Another example, of relevance here, corresponds to a vortex in one component aligned with a vortex of opposite circulation in the second component. This texture may be thought of as a lower dimensional counterpart to the ``spin vortex'' that arises spontaneously in quenched ${}^{87}$Rb spinor condensates \cite{SadHi06, *SaiKa06}.
Additional textures in 2D geometries include baby skyrmions \cite{PieSc95}, merons \cite{CalDa77}, and planar spin textures \cite{Mer79}, whose study has led to insights in reduced-dimensional superfluid or ferromagnetic systems, such as in the case of the quantum Hall effect \cite{AndMa75}.
We describe a mechanism of lattice-texture formation that extends to three or more arbitrarily located spatially-separated spinor BEC pieces with arbitrary initial coherent phases. Although related to the Kibble-Zurek (KZ) mechanism \cite{Kib76}, we focus on the vortex generation process in limits where linear interference applies. A fuller description of the KZ mechanism might involve additional description of subsequent vortex dynamics as part of an unspecified thermalization process.

In the following, we present simulations of a proposed experiment to generate the lattice texture via the interference of a nonrotating pseudospin-$\tfrac12$ BEC, initially divided into three pieces. Although it is possible to fully control the piece phases and resulting textures, we present a simplified experiment in which the phases are fixed. A planar hexagonal lattice-texture results, with a motif composed of three textures: one HQV of each sign and a spin-2 texture. We classify these textures by their topologies, and present a lattice model whose time-dependent growth is related to the initial BEC geometry.

%-----------------------------------------------------------------------------------------------------

We numerically model a two-level ${}^{87}$Rb BEC system with $\ket{F=1, m_F=-1} \! \equiv \! \ket{1}$ and $\ket{F=2, m_F=+1} \! \equiv \! \ket{2}$ using a mean-field approach. These hyperfine states and their coupling behavior have been well-studied \cite{MatHa98, MerMe07} and are convenient for the study of pseudospin-$\tfrac12$ condensates.
We performed 2D simulations, corresponding to pancake condensates. This geometry has the advantage that axial ballistic expansion is rapid, due to the initial tight axial confinement. Nonlinear effects, which might otherwise lead to distortion of any lattice and curvature or reconnection of vortex lines, are consequently short-lived.
The order parameter field of a single-species BEC whose atoms occupy two internal hyperfine levels is a position-dependent two-component pseudospinor $\Psi(\gvec{r})=\big(\Psi_1(\gvec{r}), \Psi_2(\gvec{r})\big)$, where $\gvec{r}$ is a position vector \cite{Ho98,*OhmMa98}.
The dynamical evolution of the BEC is governed by two coupled Gross-Pitaevskii Equations (GPEs):
\begin{equation}
\label{eqn:coupled_2DGPEs}
i \hbar \dfrac{\partial \Psi_i}{\partial t} = \left( -\frac{\hbar^2}{2m}\nabla^2_{\!\perp} + V_i + \Gamma(t) \!\! \sum_{j=1,2} \! U_{ij} \left| \Psi_j \right|^2 \right) \Psi_i, \ i=1,2,
\end{equation}
where $\Psi_i$ is the 2D order parameter of component $\ket{i}$ and $\nabla^2_{\!\perp}$ is a 2D Laplacian. The self-interaction parameter $U_{ij}=4\pi\hbar^2 a_{ij}/m$ depends on the intra- and inter-component $s$-wave scattering lengths $a_{ij}$ and the mass $m$ of an atom of the condensed species; there are three independent scattering lengths, since $a_{12}=a_{21}$. A time-dependent factor $\Gamma(t)$ results from the reduction from 3D to 2D, as described below.
The normalization condition is $\sum_i \int |\Psi_i|^2 \, d\gvec{r} = N$, where the total number of atoms $N$ is preserved independently of the internal spin state.

We use the scattering lengths reported by \citet{MerMe07}. The state-dependent scattering lengths are $a_{11}\!=\!100.40\,a_0$, $a_{22}\!=\!95.00\,a_0$, and  $a_{21}\!\equiv\!a_{12}\!=\!97.66\,a_0$, where $a_0$ is the Bohr radius. The condensate contains $N = 50,000\,$atoms of ${}^{87}$Rb, each of mass $m = 1.4188\times 10^{-25}\,$kg. This relatively small population size was chosen to minimize nonlinear perturbative effects and maximize the regularity of the lattice.

Initially the BEC is tightly trapped in the axial direction, producing a pancake geometry, and further divided into three pieces by trapping within three transverse Gaussian potential wells, such as would be formed by three red-detuned lasers.
The initial equilibrium condensate profile is established with all atoms in $\ket{1}$ by numerically evolving Eq.~\eqref{eqn:coupled_2DGPEs} through imaginary time.
In simulations, the numerical procedure establishes a uniform phase $(\varphi_1\!=\!\varphi_2\!=\!\varphi_3)$ for the wavepackets, which fixes the translation of the resulting lattice texture. However, in typical experiments the initial wavepacket phases are random, resulting in the lattice being randomly translated. Therefore, in an experiment, both components must be imaged simultaneously to correctly reconstruct the lattice texture structure. This might be performed by dual state imaging \cite{AndTi09} or by nondestructive phase contrast imaging \cite{TurDo05}.

Having established the initial profile, we immediately remove all trap potentials $V_i$ at $t=0$.
We account for ballistic expansion along the axial direction by assuming each piece to be a non-interacting gas with a Gaussian axial profile. Upon trap removal, the resulting rapid axial expansion and associated rapid decrease of the nonlinear coupling term are modeled by the time-dependent factor $\Gamma(t) = \left[m\omega/(2\pi\hbar) \right]^{1/2} ( 1+t^2 {\omega}^2 )^{-1/2}$, where $\omega$ is the angular frequency of the axial harmonic trap initially confining the condensate.
%Because all atoms begin in state $\ket{1}$ 
We specify $\omega\!=\!500\,\mathrm{rad\:s}^{-1}$, corresponding to a pancake geometry.

\begin{figure}[bt]
%     \centering
%     \includegraphics[width=58mm]{fig2}
     \includegraphics{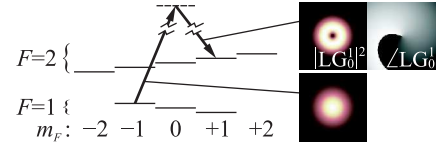}
  \caption{(color online) ${}^{87}$Rb hyperfine levels $\sket{F=1, m_F=-1}\equiv\sket{1}$ and $\sket{F=2, m_F=+1}\equiv\sket{2}$ are optically coupled with Gaussian and Laguerre-Gauss (LG) lasers, whose intensity profiles are shown. The $\textrm{LG}^{\ell=1}_{p=0}$ phase is also shown winding from $-\pi$ (black) to $+\pi$ (white).}
  \label{fig:overlaid_lattices}
\end{figure}

We simulate the application of a two-photon $\pi/2$ pulse to excite half the atoms to $\ket{2}$ [Fig.~\ref{fig:overlaid_lattices}].
This models an optical process involving two lasers, coupling $\ket{1}$ and $\ket{2}$ via an intermediate level. The $\pi/2$ pulse may be applied either immediately before or after trap removal.
The overall translation of each VA lattice depends only on the relative phases of the initial wavepackets \cite{RubPa08}. Although intra-component phases are uncontrolled in typical experiments, full control over the inter-component phase may be realized by spatially-localized Raman beam pairs focused on each wavepacket.
Atoms excited to $\ket{2}$ thereby acquire relative phases that produce a lattice which is predictably aligned with the lattice in $\ket{1}$, enabling the production of a continuum of related lattice-textures. In our simplified example, we instead employ a Laguerre-Gauss $\textrm{LG}^{\ell=1}_{p=0}$ mode in one of the coupling beams to establish wavepacket phases $\varphi_1+\Delta\varphi-2\pi/3$, $\varphi_2+\Delta\varphi$, and $\varphi_3+\Delta\varphi+2\pi/3$ for atoms in $\ket{2}$ \cite{AndRy06}. The phase offset $\Delta\varphi$ has no effect on lattice translation \cite{RubPa08}.
The LG beam wavefront confers phase gradients on each wavepacket in $\ket{2}$. Any effects of this nonuniformity are minimized by virtue of the initial tight transverse confinement of the wavepackets.

\begin{figure}[bt]
%     \centering
%     \includegraphics[width=86mm]{fig3}
    \includegraphics{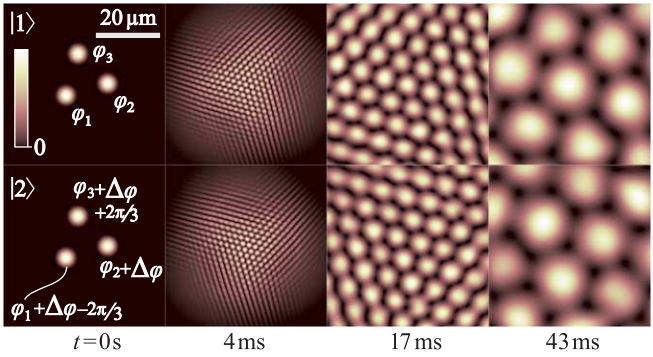}
  \caption{(color online) Numerical simulations of a two-component three-piece BEC generate vortex-antivortex lattices as the wavepackets expand and interfere. For the initial phases shown, the lattices align as shown in Fig.~\ref{fig:spacings}(a).}
  \label{fig:components}
\end{figure}

The BEC then evolves governed by Eq.~\eqref{eqn:coupled_2DGPEs}, resulting in the axially-projected probability densities $|\Psi_i|^2$ shown in Fig.~\ref{fig:components}.
A honeycomb VA lattice is formed in each component, resulting from interference of the expanding wavepackets. The equal initial phases $(\varphi_1\!=\!\varphi_2\!=\!\varphi_3)$ produce the particular lattice translations shown. The final frame of this figure shows the individual components $|\Psi_i|^2$ at $t=43\,$ms, corresponding to the lattice texture in Fig.~\ref{fig:carpet_with_bloch}.

\begin{figure}[b!]
%    \centering
%    \includegraphics[width=66mm]{fig4}
    \includegraphics{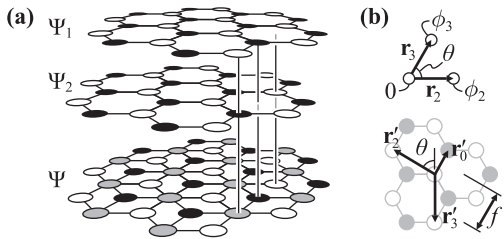}
  \caption{(a)~Lattices containing vortices (white circles) and antivortices (black circles) form in components $\Psi_1$ and $\Psi_2$. Three combinations form different textures in $\Psi\equiv(\Psi_1,\Psi_2)$. Gray circles indicate an aligned vortex and antivortex in the two components. (b)~Wavepackets with centers given by $\gvec{r}_1\equiv\gvec{0}$, $\gvec{r}_2$, $\gvec{r}_3$ have relative phases $0, \phi_2, \phi_3$, respectively. The resulting vortex-antivortex lattice basis vectors are $\gvec{r}'_2$ and $\gvec{r}'_3$, with a motif containing a vortex and antivortex separated by $\gvec{r}'_0$. Fringe spacing $f$ is a convenient measure of unit cell size.}
  \label{fig:spacings}
\end{figure}

\begin{figure}[bt]
%    \centering
%    \includegraphics[width=86mm]{fig5}
    \includegraphics{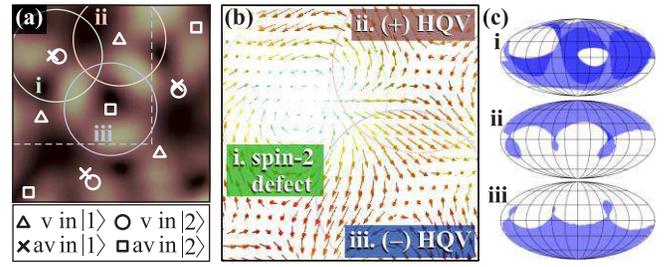}
  \caption{(color online) Three texture types within the motif. (a) Vortex (v) and antivortex (av) cores are overlaid on the combined BEC components. (b) Bloch vectors are shown normalized to the combined probability density. (c) Order-parameter-space coverage for the three circular regions, where shading indicates either one or two overlapping projections.}
  \label{fig:defect_ids}
\end{figure}

We now consider the three textures that combine in the motif of Fig.~\ref{fig:carpet_with_bloch}, corresponding to the different alignments of vortices shown schematically in Fig.~\ref{fig:spacings}(a).
We examine these textures in more detail in Fig.~\ref{fig:defect_ids}, using simulation results at $t=69\,$ms.
The locations of vortices and antivortices in each component are identified in Fig.~\ref{fig:defect_ids}(a).
Because textures extend over a local neighborhood of the field, we specify three circular bounded regions, labeled (i), (ii), and (iii), centered on the vortex cores.
Each boundary is chosen to lie approximately halfway to the nearest neighboring vortex core.
The same boundaries are shown in Fig.~\ref{fig:defect_ids}(b), which plots the field of Bloch vectors with lengths scaled by local density, and identifies the three texture types. In crystallographic terms, the three textures constitute a motif, within a hexagonal lattice.

In the two-component BEC system described here, the Bloch vectors within each bounded region of the physical space are projected onto a unit Bloch-sphere to reveal the texture topology. Maps of this order-parameter space are obtained by grouping adjacent unit Bloch vectors within the indicated circular regions, and coloring the patch of the sphere surface onto which they project. The maps are shown in Fig.~\ref{fig:defect_ids}(c) using a flattened equal-area sphere projection.

Region (i) in Fig.~\ref{fig:defect_ids} corresponds to a spin-2 texture with a zero net mass current, associated with the two constituent unit charge vortices of opposite sign that reside in different components.
In traversing a small closed contour about the vortex core in Fig.~\ref{fig:defect_ids}(b) once, the vector projections wrap the Bloch sphere twice near the equator; the winding number is 2. As contours of larger radius are traversed, the projections deviate from the equator toward the poles of the Bloch sphere. This texture is similar to spin textures described elsewhere in a system of 2D spins \cite{Mer79} and in three-component spin textures \cite{SaiKa06}.
The two prominent holes visible in the coverage map [Fig.~\ref{fig:defect_ids}(c)(i)] are due to the influence of neighboring textures.
These holes continue to shrink as the lattice expands. Coverage of polar map-regions occurs when matter in one component coincides with a complete absence of matter in the other component, i.e., at vortex cores.
Since the region (i) boundary is chosen to exclude neighboring vortex cores, the polar regions are not covered.
The vortices associated with the spin-2 texture are slightly misaligned, warping the texture.

Regions (ii) and (iii) in Fig.~\ref{fig:defect_ids} are centered on a vortex in one or other of the components. These correspond to HQVs. The sign of the HQV manifests as coverage of the corresponding hemisphere of the Bloch sphere [Fig.~\ref{fig:defect_ids}(c)(ii)--(iii)]. A threefold symmetry is apparent near the equator, due to the hexagonal symmetry of the lattice texture. As the lattice expands, the projections on the boundary approach the equator. By identifying all points on the equator with a single value, the order-parameter space is restricted to one hemisphere. The HQVs are then topological defects, since the hemispherical order parameter spaces will be covered once by similar boundaries to those shown.
An HQV has wrapping number 1 for a hemispherical order parameter space, associated with its constituent unit charge vortex. It covers a $2\pi$ solid angle and is thus a $2\pi$-defect whose sign depends on both the vortex sign and its resident component.
Similarly, the boundary projections of the spin-2 texture asymptotically approach lines of longitude on the Bloch sphere.
If these linear boundaries are ``healed,'' the wrapping number of the sphere is 2 and we may identify the texture as an $8\pi$-defect.

Having discussed the topology of the isolated textures, we now describe the lattice texture. The isolated textures expand along with the evolving lattice according to previously described expressions for vortex core locations in a model of interfering Gaussian wavepackets \cite{RubPa08}. For vortices to be produced by interference, the contributions from each expanding wavepacket must be approximately equal. This establishes a spatiotemporal condition, which limits the lattice extent and describes its growth by vortex formation within an expanding circular envelope \cite{RubPa08}. By evaluating the locations of two adjacent vortices of the same sign within this envelope, we find that the lattice basis vectors [Fig.~\ref{fig:spacings}(b)] have lengths
\begin{equation}
  |\gvec{r}'_2| = \pi / (r_3 \alpha \sin\theta), \ \ \ \ 
  |\gvec{r}'_3| = \pi / (r_2 \alpha \sin\theta),
\end{equation}
where $r_j\equiv |\gvec{r}_j|$ are the source spacings, and $\theta$ is the interior angle at the origin of the triangle of wavepackets [Fig.~\ref{fig:spacings}(b)]. These expressions contain a time-dependent lattice scaling factor $\alpha = m \hbar t/[2(\hbar t)^2+2m^2(\hbar/\Delta p)^4]$, which assumes that the wavepackets share a single initial momentum uncertainty $\Delta p$, and are therefore all of the same initial size.
The distance $f$ in Fig.~\ref{fig:spacings}(b) corresponds to the bright fringe spacing, measured experimentally by \citet{HenRy09}. For $r \equiv r_2 = r_3$ and $\theta=\pi/3$, $f=|\gvec{r}'_3| \sqrt{3}/2 = \pi / (r \alpha)$.
The spacing of adjacent vortices and antivortices in the VA lattice is
\begin{equation}
 \left|\gvec{r}'_0 \right| = \pi \bigl({r_2}^2 + 2 r_2 r_3 \cos\theta + {r_3}^2 \bigr)^{1/2} \ \big/ \ (3 r_2 r_3 \alpha \sin\theta).
\end{equation}
For $r \equiv r_2 = r_3$ and $\theta=\pi/3$, this reduces to $|\gvec{r}'_0 | = 2\pi/(3 r \alpha)$.

%-----------------------------------------------------------------------------------------------------

We have shown with numerical simulations that a lattice-texture forms in a nonrotated two-component BEC, initially separated into three pieces, and subsequently allowed to expand and interfere. We created an expanding hexagonal lattice-texture by employing spatiotemporal wavefunction engineering to determine the specific relative phases of the initial pieces and arrange them at the corners of an equilateral triangle. By employing a Laguerre-Gauss beam in the phase-engineering task, the lattice was created with a motif composed of two half-quantum vortices of opposite signs and one spin-2 texture, both being examples of direct relevance in related condensed matter systems. More generally, the method presented provides a means for the deterministic production of a continuum of related lattice-textures.

\acknowledgments{G.R. thanks T.P. Simula, K. Helmerson, L.D. Turner, R.P. Anderson, and E.J. Mueller for helpful discussions, and acknowledges Australian government support.}

%-----------------------------------------------------------------------------------------------------
%\bibliography{vortices}
%merlin.mbs 2010-03-15 4.21a (PWD, AO, DPC)

\end{document}